\documentclass[%
 reprint,
 amsmath,amssymb,
 aps,
 pra,
 longbibliography,
 showkeys,
]{revtex4-2}

\usepackage{placeins}
\usepackage{siunitx}
\usepackage{svg}
\usepackage{graphicx}
\usepackage{dcolumn}
\usepackage{bm}

\usepackage[english]{babel}

\newcommand{\E}{\mathbb{E}}
\DeclareMathOperator{\Var}{Var}
\DeclareMathOperator{\CoV}{CV}

\begin{document}

\preprint{APS/123-QED}

\title{Cell size distributions in lineages}

\author{Kaan \"Ocal}
\email{kaan.ocal@unimelb.edu.au}
\author{Michael P.H. Stumpf}%
\affiliation{%
School of BioSciences \& School of Mathematics and Statistics, University of Melbourne, \\
Parkville, Victoria 3052, Australia
}%

\date{\today}

\begin{abstract}
Cells actively regulate their size during the cell cycle to maintain volume homeostasis across generations. While various mathematical models of cell size regulation have been proposed to explain how this is achieved, relating these models to experimentally observed cell size distributions has proved challenging. In this paper we present a simple formula for the cell size distribution in lineages as observed in e.g.~a mother machine, and provide a new derivation for the corresponding result in populations, assuming exponential cell growth. Our results are independent of the underlying cell size control mechanism and explain the characteristic shape underlying experimentally observed cell size distributions. We furthermore derive universal moment identities for these distributions, and show that our predictions agree well with experimental measurements of \textit{E.~coli} cells, both on the distribution and the moment level. 
\end{abstract}

\keywords{cell division, stochastic modeling, renewal theory}

\maketitle


\section{Introduction}

Growing cells time their division to regulate their size across generations, a phenomenon known as cell size homeostasis. A range of models have been proposed to explain this phenomenon, most famously the sizer/adder/timer triptych \cite{amir_cell_2014} and its extensions. While live cell tracking can be used to probe cell cycle dynamics and its relation to cell size \cite{tanouchi_long-term_2017}, most biological experiments rely on snapshot measurements that do not capture such dynamical information. Since cell size is a critical actor affecting most cellular processes, including metabolism and gene expression \cite{marshall_what_2012,padovan-merhar_single_2015,foreman_mammalian_2020}, understanding how cell sizes behave in snapshot measurements is crucial for quantitative modeling of such phenomena. 

For forward lineages such as those observed in a mother machine, cell size distributions have been computed in special cases \cite{marantan_stochastic_2016,jia_cell_2021,genthon_analytical_2022}, but a general solution seems to be missing from the literature. The case for populations under the assumption of perfectly symmetric division was recently treated in \cite{hein_asymptotic_2024}. In this paper we generalize these results and show that for the biologically relevant case of exponential cell size growth, cell size distributions in lineage (and population) experiments can be computed directly from the birth or division size distributions. These in turn can be derived or approximated for many models of cell size dynamics such as the sizer and adder models. Our results hold for general models of cell size regulation with multi-generational memory and stochastic cell growth, and do not require knowledge of the exact mechanism behind cell size regulation. This provides a mathematical explanation for the distribution shapes observed in lineage experiments \cite{jia_cell_2021}, which often resemble that of a smeared-out log-uniform distribution. We also derive simple universal moment identities that relate moments of the lineage distribution to those of cells at birth or division. Analyzing a dataset of \textit{Escherichia coli} growth over generations, we find good qualitative and quantitative agreement with experimental measurements. 

\begin{figure*}
    \centering
    \includegraphics{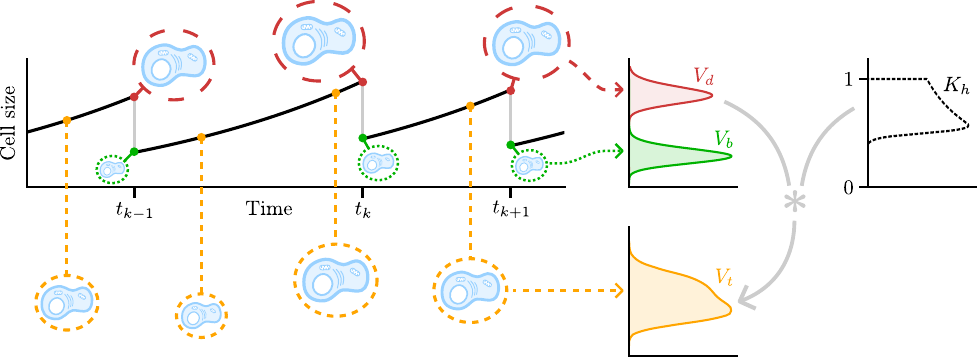}
    \caption{\label{fig:schema}Measuring cell size in a lineage experiment. Cell size grows exponentially and is partitioned into two at each division. The size $V_d$ of the mother (resp.~the tracked daughter) at each division event follows the division size (resp.~birth size) distribution. Experimental measurements at a fixed time follow the lineage distribution, which can be represented as a multiplicative convolution of the division size distribution with a division kernel $K_h$. Assuming ergodicity, measuring one lineage at regular intervals is equivalent to measuring independent lineages at a fixed time - this is not always given, see \cite{jafarpour_cell_2019}.}
\end{figure*}

\section{Analytical Results}

We consider a general model of cell size regulation illustrated in Fig.~\ref{fig:schema}. A cell born with size $V_b$ grows until it reaches a division size $V_d$, after which it divides into two daughter cells with sizes $h V_d$ and $(1 - h) V_d$, where $h \in [0,1]$ is the volume fraction inherited by the first cell. A \emph{forward} lineage is obtained by picking one of the daughter cells to track at each stage, which itself proceeds to grow and divides in the same manner.

We assume that cells grow exponentially with fixed growth rate $\gamma$; an extension to stochastic growth rates will be discussed in Appendix~\ref{app:stoch}. We assume that the division size of a cell is stochastic and depends  on its birth size via a transition kernel
\begin{equation}
    k(w, v) = p(V_{d} = w \, | \, V_{b} = v). \label{eq:volume_transition_kernel}
\end{equation}

\noindent The sizer, adder and timer models described in \cite{amir_cell_2014} are all of this form, as is any combination thereof (cf.~\cite{varsano_probing_2017}), but our setup applies equally well to mechanistic models based on accumulation thresholds for proteins \cite{schmoller_dilution_2015,ghusinga_mechanistic_2016,keifenheim_size-dependent_2017}. Our derivation will remain valid for multigenerational cell size control mechanisms where the division size of a cell depends on previous birth and division sizes \cite{elgamel_multigenerational_2023}.

For simplicity we assume that the volume fraction $h \in (0,1)$ inherited by the tracked daughter cell is independent of the division size and follows a fixed distribution $p_h(h)$. We do not restrict ourselves to symmetric division where $\E[h] = 1/2$, and we explicitly allow biased tracking protocols where the larger (or smaller) daughter is tracked over time. The case $p_h(h) = \delta(h - 1/2)$ represents perfectly symmetric division. 

The above model describes lineage dynamics as a Markov renewal process, where the birth sizes, $V_{b,1}, V_{b,2}, \ldots$, constitute a Markov chain. Our final assumption is that cell size is regulated, i.e.~the Markov chain defined by the birth sizes $V_{b,k}$ has a steady-state distribution $p_b(V)$. As shown in \cite{thomas_analysis_2018}, this distribution must satisfy 
\begin{equation}
    p_b(V) = \int_0^\infty dV_b \int_0^1 \frac{dh} h \, p_b(V_b) \, p_h(h) \, k(V/h, V_b). \label{eq:volume_stationarity}
\end{equation}

\noindent If we sample a cell with birth size \mbox{$V_b \sim p_b(V_b)$}, a corresponding division size \mbox{$V_d \sim k(V_d, V_b)$} and a fraction $h \sim p_h(h)$, the birth size of the tracked daughter, which is $h \, V_d$, again has distribution $p_b$. This rules out the timer model \cite{amir_cell_2014}, which features asymptotically diverging cell sizes contrary to biological observations. From $p_b(V)$ and Eq.~(\ref{eq:volume_transition_kernel}) we also obtain the division size distribution $p_d(V)$.

We are interested in computing the lineage distribution
\begin{equation}
    p_l(V) = \lim_{T \rightarrow \infty} T^{-1} \int_0^T dt \, p(V_t = V), \label{eq:lineage_dist_orig} 
\end{equation}

\noindent where $V_t$ is the size of the tracked cell at time $t$. This parallels the case in renewal theory, but with waiting times that are not generally independent. Markov renewal theory \cite{cinlar_markov_1969} tells us that the distribution of the birth size $V_b$ and division size $V_d$ of a cell sampled from a lineage tends asymptotically to
\begin{equation}
    p_l(V_b, V_d) = Z_l^{-1} p_b(V_b) \, k(V_d, V_b) \, \tau(V_d, V_b), \label{eq:lineage_dist}
\end{equation}

\noindent where $Z_l$ is a normalization constant and
\begin{equation}
    \tau(V_d, V_b) = \gamma^{-1} \left(\log(V_d) - \log(V_b)\right) \label{eq:tau}
\end{equation}

\noindent is the lifetime of a cell growing from size $V_b$ to $V_d$. Now observing that $h \, V_d$ and $V_b$ have the same marginal distribution in our original formulation, Eq.~(\ref{eq:volume_stationarity}), we use the product rule for logarithms to obtain
\begin{equation}
    Z_l = \E[\tau(V_d, V_b)] = -\gamma^{-1} \E\left[\log(h)\right]. \label{eq:Z_lineage}
\end{equation}

\noindent Note that Eq.~(\ref{eq:lineage_dist}) implies that the birth size distribution for the current cell in a lineage differs from $p_b(V)$, an instance of the inspection paradox. 

The size of the currently observed cell can be represented as
\begin{equation}
    V_t = V_b \, e^{\gamma a},
\end{equation}

\noindent where $a$ is the age of the cell and $V_b$ its birth size. According to Markov renewal theory, the age of a cell sampled in the stationary regime is uniformly distributed on $[0, \tau]$, that is, we can write $a = \theta \tau$ where $\theta$ is uniformly distributed on $[0,1]$ and independent of $\tau$. Multiplying Eq.~(\ref{eq:lineage_dist}) by $V_t^\alpha = V_b^\alpha e^{\gamma \alpha \theta \tau}$ and integrating we obtain the following formula for the moments of the cell size distribution in a lineage,
\begin{equation}
    \E_l[V_t^\alpha] = \frac{\E[V_d^\alpha] - \E[V_b^\alpha]}{\alpha\, \E[\log(h^{-1})]}.
\end{equation}

\noindent This is valid for any $\alpha \neq 0$. Here and in the sequel, $\E_l$ denotes expectations with respect to the lineage distribution, Eq.~(\ref{eq:lineage_dist_orig}). We can again use the fact that $h \, V_d$ and $V_b$ have the same marginal distribution to write this as
\begin{equation}
    \E_l[V_t^\alpha] = \frac{1-\E[h^\alpha]}{\alpha \, \E[\log(h^{-1})]} \, \E[V_d^\alpha]. \label{eq:moment_alpha}
\end{equation}

\noindent This expresses the Laplace transform of the lineage distribution over $\log(V_t)$ as a product of the Laplace transform of $\log(V_d)$ and a prefactor depending on $p_h$. Recognizing the pre-factor as another Laplace transform, we see that $\log(V_t)$ can be expressed as the independent sum of two random variables, or upon exponentiation that
\begin{equation}
    V_t\,  \mathrel{\mathop{\sim}\limits^d}\, V_d \, z^\theta, \label{eq:result}
\end{equation}

\noindent where $\mathrel{\mathop{\sim}\limits^d}$ denotes equality in distribution. Here the auxiliary variables $V_d$, $\theta$ and $z$ are distributed according to
\begin{eqnarray}
    V_d \sim p_d(V_d),\; \theta \sim \mathcal{U}(0, 1),\; p(z) &= \displaystyle \frac{p_h(z) \log(z)}{\E[\log(h)]} \label{eq:generative}.
\end{eqnarray}
\noindent It is important to note that in this context, $V_d$ does not represent the division size of the currently sampled cell, or its ancestor: these will in general follow different distributions due to the inspection paradox. Eq.~(\ref{eq:result}) expresses the probability distribution of $V_t$ as a multiplicative convolution of $p_d$ with a kernel $K_h$, the probability density function of $z^\theta$.

From Eq.~(\ref{eq:result}) we can derive universal moment identities that relate empirically observed cell size statistics to those of the division size distribution, such as
\begin{eqnarray}
    \E_l[\log(V_t)] &=& \E[\log(V_d)] + \frac 1 2 \frac{\E[\log(h)^2]}{\E[\log(h)]}, \label{eq:mean_logV} \\
    \E_l[\log(V_t)^2] &=& \E[\log(V_d)^2] + \frac 1 3 \frac{\E[\log(h)^3]}{\E[\log(h)]}, \\
    \E_l[V_t] &=& \E[V_d] \left(\frac {1 - \E[h]}{\E[\log(h^{-1})]}\right), \\
    \E_l[V_t^2] &=& \E[V_d^2] \left(\frac {1 - \E[h^2]}{2 \, \E[\log(h^{-1})]}\right), \label{eq:var_V}
\end{eqnarray}

\noindent generalizing some results in \cite{marantan_stochastic_2016}. Here, of course, the latter two equations are special cases of Eq.~(\ref{eq:moment_alpha}). If $h$ follows e.g.~a Beta or lognormal distribution, the relevant moments of $h$ can be computed explicitly. It can be verified that Eq.~(\ref{eq:result}) is equivalent to
\begin{equation}
    p_l(V) = \displaystyle \frac {\int_0^1 p_h(h) F_d(V/h) \, dh - F_d(V)} {V \,\E[\log(h^{-1})]}, \label{eq:alternative}
\end{equation}

\noindent where $F_d$ is the cumulative distribution function of the division size distribution $p_d$. The integral is of course equal to $F_b(V)$.

In the special case of perfectly symmetric division, we can use the fact that $V_{d,k} = 2\,V_{b,k+1}$ to relate the birth and division size distributions and represent $V_t$ as
\begin{equation}
    V_t \,  \mathrel{\mathop{\sim}\limits^d}\, V_b \, 2^\theta, \label{eq:result_perfect_division}
\end{equation}

\noindent where this time
\begin{equation}
    V_b \sim p_b(V_b),\quad \theta \sim \mathcal{U}(0, 1).
\end{equation}

\noindent Note that we are using the birth size distribution here. The moment identities in Eqs.~(\ref{eq:mean_logV})--(\ref{eq:var_V}) simplify correspondingly, and Eq.~(\ref{eq:alternative}) becomes
\begin{equation}
    p_l(V) = \frac {F_b(V) - F_b(V/2)} {V \log(2)}, \label{eq:alternative_sym}
\end{equation}

\noindent where $F_b$ is the cumulative distribution function of the birth size distribution $p_b$. This is the forward lineage version of the result established in \cite{hein_asymptotic_2024}.

We observe that Eq.~(\ref{eq:result}) is a purely formal consequence of Eq.~(\ref{eq:lineage_dist}). Our derivation does not hinge on the fact that $V_d$ only depends on $V_b$, and only relies on the joint distribution of $V_b$ and $V_d$. This allows us to extend our calculations to models of cell size control with multi-generational memory, where $V_{d,k}$ depends on $V_{b,k}$, but also on  $V_{d,k-1}$, $V_{b,k-1}$, $V_{d,k-2}$ etc., as long as we assume that birth sizes reach a stationary distribution. This will be the case if multi-generational memory is finite, i.e.~birth sizes are defined by higher-order Markov chains, or if correlations decay fast enough. 

The fact that the growth rate $\gamma$ does not appear in Eq.~(\ref{eq:result}) suggests that the equation should not be strongly affected by variations in the growth rate. Indeed, following \cite{hein_asymptotic_2024} we show in Appendix~\ref{app:stoch} that Eq.~(\ref{eq:result}) is valid in the case of stochastic growth rates, as long as growth rate fluctuations are independent of cell size. Appendix~\ref{app:biphasic} treats the case of biphasic growth, where growth rates fluctuate systematically as the cell switches between two different growth phases. Finally, in Appendix~\ref{app:pop} we derive the corresponding distribution in the population case, slightly generalizing the corresponding result in [9].
 
\section{Numerical Results}

\begin{figure*}
    \includegraphics{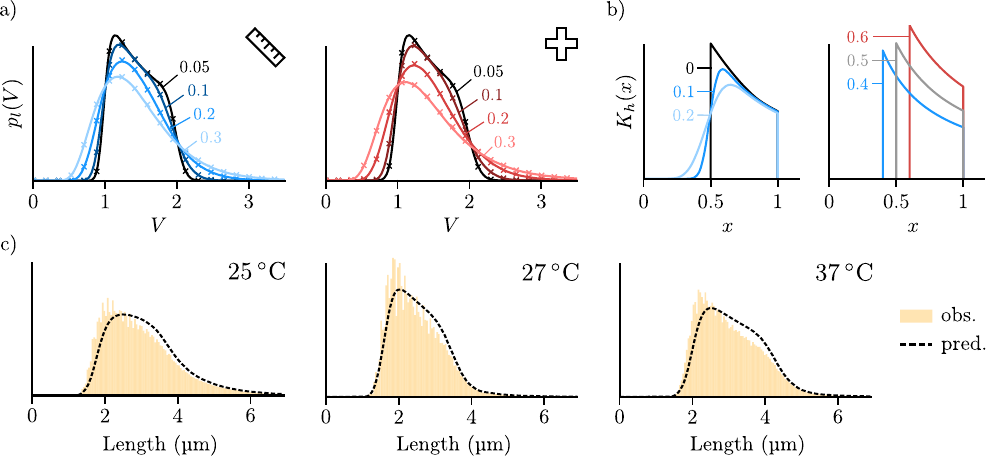}
    \caption{\label{fig:numerical}(a) Numerical estimates of the cell size distribution in a lineage (crosses) and analytical predictions (solid lines) for the sizer and adder models. We used symmetric division with coefficient of variation $\CoV_h = 0.05$, and additive Gaussian timing noise with the indicated standard deviations. (b) Visualization of the division kernel $K_h(x)$ defining the shape of the cell size distribution, for symmetric division with different values of $\CoV_h$ (left) and deterministic division with different values of $h$. (c) Empirically measured cell size distributions (shaded) and analytical predictions (dashed lines) in \cite{tanouchi_long-term_2017}.}
\end{figure*}

We verified Eq.~(\ref{eq:result}) with Monte Carlo experiments for the sizer and adder models in Fig.~\ref{fig:numerical}(a). As can be seen in our simulations and has been remarked in \cite{jia_cell_2021}, cell size distributions in a lineage have a stereotypical shape characterized by a fast increase in cells around a lower size threshold, followed by a slow quasi-exponential decay and subsequently a rapid decrease around a higher size threshold. This shape is best exemplified by the log-uniform distribution on the interval $[1/2, 1]$ (see Fig.~\ref{fig:numerical}(b)), which is the shape of the lineage distribution for deterministic cell size control and division. We showed in the previous section that the observed cell size distribution is the multiplicative convolution of the division size distribution with a division kernel $K_h$, examples of which are shown in Fig.~\ref{fig:numerical}(b). The division kernel is just a mixture of log-uniform distributions, which explains the characteristic shape we observe in experiments.

\setlength{\tabcolsep}{1pt}
\begin{table*}[t]
\caption{\label{tab:moments}
Empirically measured and predicted moments for the cell size distributions in \cite{tanouchi_long-term_2017}. Sizes were measured in \si{\micro\meter}.}
\begin{ruledtabular}
\begin{tabular}{c@{\hskip 12pt}ccr@{\hskip 12pt}ccr@{\hskip 12pt}ccr}
& \multicolumn{3}{c}{\SI{25}{\celsius}} & \multicolumn{3}{c}{\SI{27}{\celsius}} & \multicolumn{3}{c}{\SI{37}{\celsius}} \\
Moment & Measured & Predicted & Error & Measured & Predicted & Error & Measured & Predicted & Error \\
\hline
$\E_l[\log(V)]$ & 1.033 & 1.083 & 4.9\% & 0.865 & 0.902 & 4.3\% &  1.067 & 1.116 & 4.6\% \\
$\Var_l(\log(V))$ & 0.110 & 0.102 & 7.7\% & 0.0781 & 0.0754 & 3.5\% & 0.0777 & 0.0702 & 9.7\% \\
$\E_l[V]$ & 2.976 & 3.11 & 4.6\% & 2.472 & 2.56 & 3.6\% & 3.02 & 3.16 & 4.6\% \\
$\Var_l(V)$ & 1.152 & 1.164 & 1.0\% & 0.538 & 0.548 & 1.7\% & 0.813 & 0.792 & 2.7\% \\
\end{tabular}
\end{ruledtabular}
\end{table*}  

We verify our predictions experimentally using growth data from \cite{tanouchi_long-term_2017}, containing the lengths of \textit{E.~coli} cells in a lineage at fixed time points as measured in a mother machine at three different temperatures: \SI{25}{\celsius}, \SI{27}{\celsius} and \SI{37}{\celsius}. As \textit{E.~coli} is typically rod-shaped, we treat length as a proxy for cell size. For each temperature we use the empirical distributions of division sizes to estimate $p_d$, and compare lengths before and after division events to estimate $p_h$. For simplicity, we approximate the latter by a Beta distribution by matching the mean and variance. We estimate $\E[h] \approx 0.45\textrm{--}0.46$, indicating a small amount of tracking bias, and $\CoV_h \approx 0.07\textrm{--}0.09$, representing a small amount of stochasticity at division. The Pearson correlation between $h$ and $V_d$ is less than $0.1$ in all three experiments, and a visual inspection of the data does not indicate any noticeable dependence between the two. We therefore use our estimates of $p_d$ and $p_h$ to compute the lineage size distribution using Eq.~(\ref{eq:result}), which approximates the empirical cell size distribution quite well as can be seen in Fig.~\ref{fig:numerical}(c). 
As shown in Table~\ref{tab:moments}, the measured moments agree quantitatively with the values predicted using the moment identities Eqs.~(\ref{eq:mean_logV})--(\ref{eq:var_V}).

We observe a small, but consistent bias in the fits, with smaller cells occuring more often than predicted by our model. This can be explained by the fact that the cells exhibit a period of slow growth immediately after division, elongating at a much slower rate than usual. Thus we are more likely to observe cells around their birth size compared to our simple exponential model. To better capture this, we can extend our model by assuming that each cell undergoes two exponential growth phases at different rates, see Appendix~\ref{app:biphasic}. Similar biphasic behavior can be predicted assuming that cell growth changes after genome duplication and was empirically observed in \textit{Bacillus~subtilis} \cite{nordholt_biphasic_2020}.


\section{Discussion}

We have derived a general identity, Eq.~(\ref{eq:result}), for cell size distributions observed in lineage data, assuming exponential cell growth. Our approach is applicable to general mechanisms of cell size regulation and provides a model-agnostic explanation for cell size distributions commonly observed in lineage experiments, generalizing results in \cite{marantan_stochastic_2016,jia_cell_2021,genthon_analytical_2022}. As a consequence, we were able to derive universal moment identities, Eqs.~(\ref{eq:mean_logV})--(\ref{eq:var_V}), that can be tested experimentally. When applied to lineage measurements of exponentially growing \textit{E.~coli} cells, our results show good agreement with the data on both the distribution and the moment level. To our knowledge, a general relationship of the form in Eq.~(\ref{eq:result}) has not been previously established in the literature. Our approach can be applied to derive population statistics with slight modifications, and we show in Appendix~\ref{app:pop} how our methodology connects with the recently derived equivalent of Eq.~(\ref{eq:result}) for growing populations in \cite{hein_asymptotic_2024}.

The simplicity of Eqs.~(\ref{eq:Z_lineage}) and (\ref{eq:result}) suggests that analytical formul\ae{} could be derived for more quantities of interest, such as the birth or division size distributions in a lineage, or the joint size and age distribution. Indeed, using Eq.~(\ref{eq:lineage_dist}), such expressions can be obtained in terms of the joint distribution $p(V_b, V_d)$ for general models of cell growth. In most cases, however, the distributions cannot be expressed purely in terms of the birth and death size distributions and will depend on the mechanism of cell size control used. Being able to predict these lineage or population quantities in general will be helpful in modeling intracellular processes such as gene expression, signaling, metabolic activity, and stress response in the presence of cell size regulation. So far theoretical studies \cite{thomas_making_2017,beentjes_exact_2020,jia_frequency_2021} often assume timer-like dynamics, which are mathematically tractable, but unable to realistically model cell size. Accurate quantitative models of size-dependent processes, including transcription and translation \cite{beentjes_exact_2020}, must therefore rely on models that incorporate cell size homeostasis.

Eq.~(\ref{eq:result}) only holds for exponential growth, which is the most common form of cell growth in biology, but not altogether universal, even in bacteria \cite{nordholt_biphasic_2020}. In \cite{nakaoka_aging_2017} it was observed that fission yeast cells plateau in size as they approach division, which results in a second ``bump'' in the observed cell size distribution corresponding to cells about to divide. As we indicate in the appendix, the ideas in this paper can be extended to more general cases by introducing multiple stages with different growth rates, paralleling the calculations in \cite{jia_characterizing_2022}, as long as growth in each phase is approximately exponential. A more serious limitation of our study is the assumption that growth rates are assumed to be uncoupled from volume, despite biological evidence pointing at some degree of correlation \cite{kohram_bacterial_2021}. We also neglect long-term changes in cell proliferation due to environmental fluctuations and aging \cite{stewart_aging_2005}, and as a result our results are not biologically realistic in the very long-time limit. 

\begin{acknowledgments}
The authors would like to thank Andrew Mugler for insightful discussions, Yong See Foo and Augustinas Sukys for critical feedback on the manuscript, and Farshid Jafarpour for pointing out the connection with \cite{hein_asymptotic_2024}. K\"O and MPHS gratefully acknowledge financial support through an ARC Laureate Fellowship to MPHS (FL220100005).
\end{acknowledgments}

\appendix

\section{Stochastic growth rates}

\label{app:stoch}

As shown in \cite{hein_asymptotic_2024}, cell size distributions are robust to growth rate fluctuations, there treated explicitly in the case where the growth rate follows an Ornstein-Uhlenbeck process. The same argument can be applied directly to our results. We extend this by verifying that the cell size distribution does not change under another noise model, where the growth rate is constant in each generation and otherwise follows an ergodic stochastic process that is independent of cell size, such as the autoregressive model proposed in \cite{levien_non-genetic_2021}. This covers the case of variable, but possibly correlated growth rates across generations, assuming that growth rates are not directly correlated with cell size.

Assume that the growth rate in each generation follows an independent ergodic process $\gamma_1, \gamma_2, \ldots$ with stationary distribution $p_\gamma(\gamma)$. Then Eq.~(\ref{eq:lineage_dist}) becomes
\begin{equation}
    p_l(V_b, V_d, \gamma) = Z_l^{-1} \, p_\gamma(\gamma) \, p_b(V) \, k(V_d, V_b) \, \tau(V_d, V_b), \label{eq:lineage_dist_gamma}
\end{equation}

\noindent where $\gamma$ is the growth rate in the current generation and the normalization constant is given by
\begin{align}
    Z_l &= -\E[\gamma^{-1}] \, \E[\log(h)].
\end{align}

\noindent Conditioning on $\gamma$, our derivation goes through as before to yield the predicted cell size distribution Eq.~(\ref{eq:result}). Since the latter does not involve $\gamma$, it remains unchanged when we marginalize out $\gamma$, which proves our claim. The above can be recast in terms of a time-change argument as in \cite{hein_asymptotic_2024} and combined with other noise models (such as Ornstein-Uhlenbeck process) discussed in \cite{levien_non-genetic_2021}.

\section{Biphasic growth}

\label{app:biphasic}

Experimental observations suggest that cell growth is not strictly exponential in most cell types \cite{tanouchi_long-term_2017,nordholt_biphasic_2020}. Here we illustrate how our results can be extended to model biphasic growth, where we assume that cells undergo two stages of exponential growth with different growth rates.  

In our extended model, a cell with birth size $V_b$ grows with rate $\gamma_1$ to a random size $V_m$ determined by a transition kernel $k_1(V_m, V_b)$, and then with rate $\gamma_2$ to size $V_d$, determined by another transition kernel $k_2(V_d, V_m)$. Upon reaching $V_d$, the cell divides, where we still assume that the tracked daughter cell inherits a random fraction $h$ of the mother's size. The resulting Markov chain defines stationary birth and division size distributions as before, together with a stationary ``midpoint'' distribution $p_m(V_m)$. 

The generalization of Eq.~(\ref{eq:lineage_dist}) to this case involves two distributions depending on the phase of the observed cell:
\begin{eqnarray}
    p_{l,1}(V_b, V_m) = p_b(V_b) \, k_1(V_m, V_b) \, \displaystyle \frac{\tau_1(V_m, V_b)}{Z_{l,1}}, \\
    p_{l,2}(V_m, V_d) = p_m(V_m) \, k_2(V_d, V_m) \, \frac{\tau_2(V_d, V_m)}{Z_{l,2}},
\end{eqnarray}

\noindent with dwelling times
\begin{equation}
    \tau_i(w, v) = \gamma_i^{-1} \left(\log(w) - \log(v)\right) \quad (i = 1,2),
\end{equation}

\noindent and corresponding normalization constants
\begin{eqnarray}
    Z_{l,1} = \E[\tau_1(V_m, V_b)] = \displaystyle \frac{\E[\log(V_m)] - \E[\log(V_b)]}{\gamma_1}, \\
    Z_{l,2} = \E[\tau_2(V_d, V_m)] = \displaystyle \frac{\E[\log(V_d)] - \E[\log(V_m)]}{\gamma_2}.
\end{eqnarray}

\noindent The probability of observing a cell in phase $i$ is given by
\begin{equation}
    p_l(i) = \frac{Z_{l,i}}{Z_{l,1} + Z_{l,2}}. \label{eq:biphasic_phase}
\end{equation}

\noindent Conditioned on the phase of the observed cell, the observed cell size distribution satisfies
\begin{eqnarray}
    \E_l[V_t^\alpha \, | \, i = 1] = \frac{\E[V_m^\alpha] - \E[V_b^\alpha]}{\alpha \gamma_1 Z_{l,1}},\\
    \E_l[V_t^\alpha \, | \, i = 2] = \frac{\E[V_d^\alpha] - \E[V_m^\alpha]}{\alpha \gamma_2 Z_{l,2}}
\end{eqnarray}

\noindent From this we recover the cell size distributions as
\begin{eqnarray}
    p_l(V \, | \, i = 1) &= \displaystyle \frac 1 {V \gamma_1 Z_{l,1}} \left(F_b(V) - F_m(V) \right), \label{eq:biphasic_cond1} \\
    p_l(V \, | \, i = 2) &= \displaystyle \frac 1 {V \gamma_2 Z_{l,2}} \left(F_m(V) - F_d(V) \right),\label{eq:biphasic_cond2}
\end{eqnarray}

\noindent where $F_b$, $F_m$ and $F_d$ are the cumulative distribution functions of $p_b$, $p_m$ and $p_d$, respectively. This generalizes Eq.~(\ref{eq:alternative}). The marginal cell size distribution $p_l(V)$ is then obtained by combining these two equations with Eq.~(\ref{eq:biphasic_phase}).

Note that the growth rates cancel out in the conditional distributions Eq.~(\ref{eq:biphasic_cond1}) and (\ref{eq:biphasic_cond2}). Their only role is in determining the average fraction of time a cell spends in each phase via Eq.~(\ref{eq:biphasic_phase}).

\section{Population results}

\label{app:pop}

Snapshot statistics in exponentially growing populations of cells differ from those in forward lineages \cite{thomas_analysis_2018}. Primarily, population snapshots skew towards younger cells, and lineages that have undergone more divisions are overrepresented in a population. In \cite{nozoe_inferring_2017}, the authors described this phenomenon by contrasting forward lineages, which we discussed above, with \emph{retrospective}, or backward lineages, which are obtained by sampling a random cell from the population and tracing its ancestry backwards in time. The population case can therefore be tackled by analyzing backward lineages and relating these to forward lineages discussed in this paper. In this section we derive the population equivalent of Eq.~(\ref{eq:lineage_dist}) to generalize the derivation for the population cell size distribution established in \cite{hein_asymptotic_2024} to the case of asymmetric or stochastic division.

We assume that cells grow and divide according to the lineage model described in the previous section, but this time both daughter cells are represented in the population. As a result, when tracing forward lineages we pick each daughter with probability $1/2$, which requires $p_h(h) = p_h(1 - h)$, and in particular that $\E[h] = 1/2$. This still covers asymmetric division as observed with budding yeast, assuming that both daughters follow the same growth process. Here we assume a fixed, deterministic growth rate $\gamma$.

In the following we will denote by $\ell_t$ a lineage up to time $t$ and by $p(\ell_t)$ its probability under our original lineage model. Assume that the initial size $V_0$ is fixed. We are interested in the population distribution $p_p(\ell_t)$ over lineages. Following \cite{nozoe_inferring_2017}, this distribution is given by
\begin{equation}
    p_p(\ell_t) = N(t)^{-1} \, p(\ell_t) \, 2^{\Delta(\ell_t)}, \label{eq:pop_weight}
\end{equation}

\noindent where $\Delta(\ell_t)$ is the number of times the lineage has divided and $N(t)$ is the expected population size at time $t$, asymptotically growing as
\begin{align}
    N(t) &= Z e^{\Lambda t}, \label{eq:pop_growtheq}
\end{align}

\noindent for some constant $Z$ and $\Lambda$ the population growth rate. As shown in \cite{lin_effects_2017}, assuming constant exponential volume growth we have $\Lambda = \gamma$. 

Let $t_1, \ldots, t_n$ be the division times of our lineage, where $n = \Delta(\ell_t)$ is the number of divisions. If $h_1, \ldots, h_n$ are the size fractions inherited at each division then we can write slightly informally
\begin{equation}
    p(\ell_t) = p(\ell_t \, | \, h_1, \ldots, h_n) \prod_{i=1}^n p_h(h_i),
\end{equation}

\noindent where the first term denotes the conditional distribution of all other variables in the lineage. The birth size of the currently alive cell is given by
\begin{equation}
    V_{b,n} = e^{\gamma t_n} \, V_0 \, h_1 \cdots h_n. \label{eq:pop_volrel}
\end{equation}

\noindent In particular, $V_0$ and the $h_i$ uniquely define the division times of a lineage. Using this identity we can write
\begin{equation}
    2^n e^{-\gamma t} p(\ell_t) = e^{-\gamma a} \frac{V_0}{V_b} p(\ell_t \, | \, h_1, \ldots, h_n) \prod_{i=1}^n 2 h_i \, p_h(h_i), \label{eq:pop_intermediate}
\end{equation}

\noindent where $a = t - t_n$ is the age of the current cell. We introduce the retrospective (backward) partition distribution 
\begin{equation}
    p_{r,h}(h) = 2h \, p_h(h), \label{eq:ph_pop}
\end{equation} 

\noindent which is normalized by our assumption that $p_h(h) = p_h(1 - h)$. If we plug Eq.~(\ref{eq:pop_intermediate}) into Eq.~(\ref{eq:pop_weight}) we obtain
\begin{equation}
    p_p(\ell_t) = Z^{-1} e^{-\gamma a} \frac{V_0}{V_b} p(\ell_t \, | \, h_1, \ldots, h_n) \prod_{i=1}^n \, p_{r,h}(h_i). \label{eq:pop_almost}
\end{equation}

\noindent Now consider the retrospective lineage distribution $p_r(\ell_t)$, where we replace $p_h$ by $p_{r,h}$ in our original model, without changing the transition kernel $k(V_d, V_b)$. We can rewrite our equation as
\begin{equation}
    p_p(\ell_t) \propto V_b^{-1} e^{-\gamma a} p_r(\ell_t), \label{eq:pop_ultimate}
\end{equation}

\noindent with proportionality constant depending on $Z$ and $V_0$. 

Eq.~(\ref{eq:pop_ultimate}) shows that typical lineages in a population follow the retrospective lineage model, which replaces $p_h$ by Eq.~(\ref{eq:ph_pop}). This results in different ancestral birth and division size distributions $p_{r,b}(V_b)$ and $p_{r,d}(V_d)$, which are the stationary distributions for the modified Markov chain. This argument provides a probabilistic derivation of the same result in \cite{thomas_single-cell_2017,genthon_analytical_2022}; we remark that this follows more generally from the optimal lineage principle in \cite{wakamoto_optimal_2012,genthon_fluctuation_2020}. Note that the forward and retrospective lineage models coincide for perfectly symmetric division where $p_h(h) = \delta(h - 1/2)$. 

We immediately arrive at the following population analogue to Eq.~(\ref{eq:lineage_dist}):
\begin{equation}
    p_p(V_b, V_d, a) = Z_p^{-1} V_b^{-1} e^{-\gamma a} p_{r,b}(V_b) \, k(V_d, V_b), \label{eq:pop_dist}
\end{equation}

\noindent where $a \leq \tau(V_b, V_d)$ and the normalization constant $Z_p$ is given by
\begin{equation}
    Z_p = \E_r[V_b^{-1}] - \E_r[V_d^{-1}] = \E_r[V_d^{-1}], \label{eq:Z_p}
\end{equation}

\noindent since $\E_r[h^{-1}] = 2$ according to Eq.~(\ref{eq:ph_pop}). 

From Eq.~(\ref{eq:pop_dist}) and Eq.~(\ref{eq:Z_p}) we obtain the following analogue of Eq.~(\ref{eq:moment_alpha}):
\begin{equation}
    \E_p[V_t^\alpha] = \frac{\E_r[V_d^{\alpha-1}]}{\E_r[V_d^{-1}]} \frac{1 - \E_r[h^{\alpha - 1}]}{\alpha - 1},
\end{equation}

\noindent for $\alpha \neq 1$. In this case the analogue to Eq.~(\ref{eq:result}) is more opaque than the result for forward lineages, but we can write the probability density function as
\begin{equation}
    p_p(V) = \displaystyle \frac{\int_0^1 p_{r,h}(h) F_{r,d}(V/h) \, dh - F_{r,d}(V)}{V^2 \, \E_r[V_d^{-1}]},
\end{equation}

\noindent which recovers the result in \cite{hein_asymptotic_2024} for the case of deterministic symmetric division, keeping in mind that the latter implies $V_d = 2V_b$.

\FloatBarrier

\bibliography{bibliography}

\end{document}